\documentclass{article}
\usepackage{graphicx,makeidx,times,a4wide}
\begin{document}


\def\ap{\textrm{'}}
\def\Biops{{\sc Biops} }
\def\Biopsn{{\sc Biops}}
\def\Oops{{\sc Oops} }
\def\Oopsn{{\sc Oops}}
\def\Aixi{{\sc Aixi} }
\def\Aixin{{\sc Aixi}}
\def\tl{{\sc Aixi}{\em (t,l)} }
\def\tln{{\sc Aixi}{\em (t,l)}}
\def\hs{{\sc Hsearch} }
\def\hsn{{\sc Hsearch}}
\def\GM{G\"{o}del Machine }
\def\gm{G\"{o}del machine }
\def\GMn{G\"{o}del Machine}
\def\gmn{G\"{o}del machine}
\newtheorem{method}{Method}[section]
\newtheorem{principle}{Principle}
\newtheorem{procedure}{Procedure}[section]
\def\odt{{\textstyle{1\over 2}}}

\title{Evolution of National Nobel Prize Shares in the 20th Century}

\date{September 2010}
\author{J\"{u}rgen Schmidhuber \\
IDSIA, Galleria 2, 6928 Manno-Lugano, Switzerland \\
University of Lugano \& SUPSI, Switzerland 
}


\maketitle

\begin{abstract}
\noindent
We analyze the evolution of cumulative national shares of Nobel Prizes since 1901,
properly taking into account that most prizes were divided among several laureates.
We rank by citizenship at the moment of the award, and by country of birth.
Surprisingly, graphs of this type have not been published before, 
even though they powerfully illustrate the century's migration patterns 
(brain drains and gains)  in the sciences and other fields.
\end{abstract}


The Nobel Foundation does not treat all Nobel Laureates equally.
While some get a full prize for their achievements, most get only 1/2, 1/3, or 1/4.
For example, some Nobel Prizes are divided among three laureates such that 
one of them gets 1/2, and the others 1/4 each.
Our data gathered from the Nobel Foundation web site {\em nobelprize.org} 
(retrieved in March 2010)
takes this into account.
For the most successful nations of the 20th century,
our figures show the temporal evolution from 1901 to 2000 (and up to 2009) 
of national Nobel Prize shares, by country of birth,
by citizenship at the time of the award, 
for each Nobel Prize type, for the total, and for the sciences in particular.

{\bf Methodology.}
For any Nobel laureate with $n$ nationalities we add
his/her award fraction (1, 1/2, 1/3, or 1/4) to the $n$
corresponding national counts.\footnote{That is, we do not multiply the 
award fraction by $1/n$, which would ensure that
all prizes of all nations add up correctly.
The graphs would hardly change though if we did, because
according to the Nobel Foundation, only
few laureates---mostly in the US and the UK---had more than one 
nationality when they got their award.}
Prizes for organizations are treated like those for individuals.
We combine prizes of laureates from Russia and the USSR.
We do not include the ``Sveriges Riksbank Prize in Economic Sciences in Memory of Alfred Nobel''
handed out since 1969 together with the original prizes.\footnote{
The prize of the Swedish bank is not an official Nobel Prize,
although the popular press often calls it that---the Nobel 
Foundation carefully sets it apart from the others.
Since it did not even exist for more than two thirds of the century, 
its inclusion would also somewhat
distort the graphs, although  the broad overall picture would not change much.}
In Figures 1-14, 
countries are ordered from upper left 
to lower right corner by
the year their first laureate received a Nobel Prize fraction.
For any year the vertical width of some nation's colored band
measures its percentage of Nobel prizes of the given type
up to that year (image height $=$ 100\%). That is, each vertical slice
of a given year compactly summarizes the information available 
to a person living in that year.
Color codes of countries are largely consistent across figures, 
but sometimes adjusted to improve contrast on grey-scale printers.

\begin{figure*}[hbt]
\centering
\includegraphics[width=\textwidth,height=\textwidth]{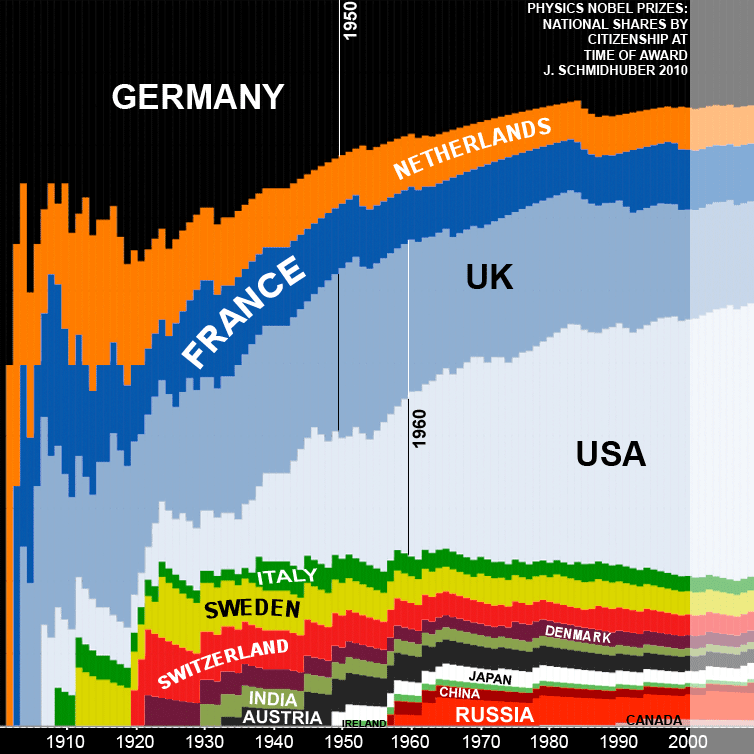}
\caption{
{\bf Physics} Nobel Prizes 1901-2009: 
Evolution of cumulative national shares {\bf by citizenship} at the time of the award.
For any year the vertical width of some nation's colored band
measures its percentage of all Nobel prizes 
up to that year (image height $=$ 100\%). 
Countries are ordered from upper left to lower right corner by
the year their first laureate received a Nobel Prize fraction.
Germany took the lead in 1901, 
the Netherlands caught up in 1902, 
France in 1903,
the UK in 1904;
Germany led alone in 1905,
the UK caught up in 1906,
France in 1908, 
then Germany led alone until 1948,
then shared the lead with the UK until 1950;
then the UK led alone until 1959 when the US caught up; 
in 1960 the US took over for good.
Note the substantial differences 
to the birth-based ranking of Figure \ref{physnat754},
reflecting brain drain from Germany 
and other countries such as Hungary to the US (and also the  UK).
}
\label{phys754}
\end{figure*}

{\bf Rankings Based on Citizenship.}
In 1956, the US started to lead the total Nobel Prize count by citizenship at the
moment of the award, taking over from Germany
(which shared the lead with the United Kingdom for one year  1904-05; Figure \ref{all754}).
The UK passed France for good in 1934, and Germany in 1974.
Considering only Nobel Prizes in the sciences (Figure \ref{sci754}), 
and ignoring high-variance fluctuations with little statistical significance in the beginning
of the century,
Germany was ahead until 1964:
until 1984 in chemistry (after a quick start by the Netherlands; see Figure \ref{chem754}), 
until 1953 in medicine (Figure \ref{med754}),
until 1950 in physics (led by the UK in the 1950s; Figure \ref{phys754}).
Since 1984, all sciences have been led by the US. 
Switzerland, US, and UK collected many
of the peace prizes  (the US has led this category
since 1929 although former leader Switzerland briefly caught up 
again in 1944; see Figure \ref{pea754}).
France led the literature count for most of the century (Figure \ref{lit754}).

\begin{figure*}[hbt]
\centering
\includegraphics[width=\textwidth,height=\textwidth]{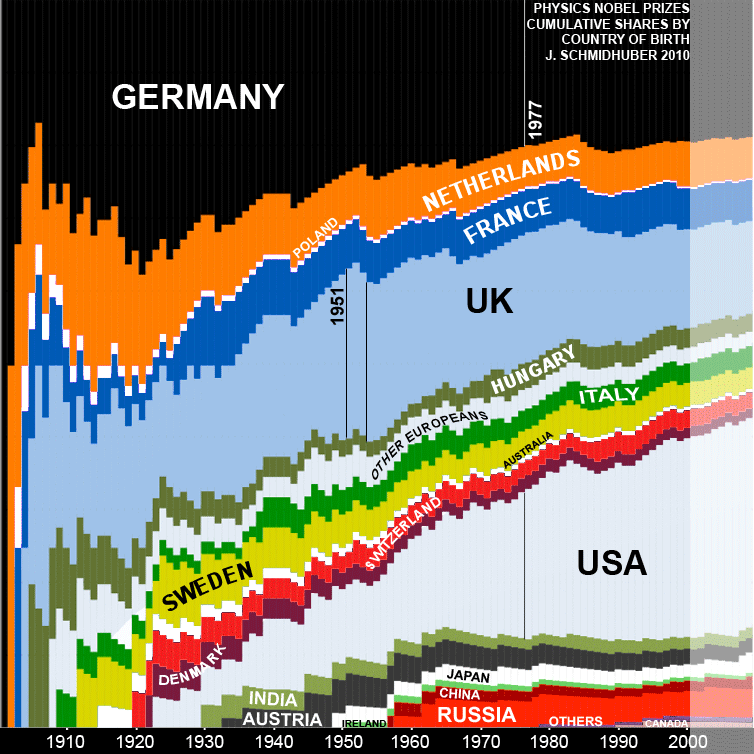}
\caption{
{\bf Physics} Nobel Prizes 1901-2009: 
Evolution of cumulative national shares {\bf by country of birth.}
Germany took the lead in 1901, 
the Netherlands caught up in 1902, 
the UK in 1904;
the UK took the lead in 1906, then
shared it in 1907 with Germany,
which then led alone until 1976 
(except for the period 1951-54 when the UK caught up for 3 years).
In 1976, German-born  shared the lead with US-born,
who took over for good in 1977.
Note the  differences 
to the citizenship-based ranking of Figure \ref{phys754},
reflecting brain drain to the US (and the UK).
}
\label{physnat754}
\end{figure*}

\begin{figure*}[hbt]
\centering
\includegraphics[width=\textwidth,height=\textwidth]{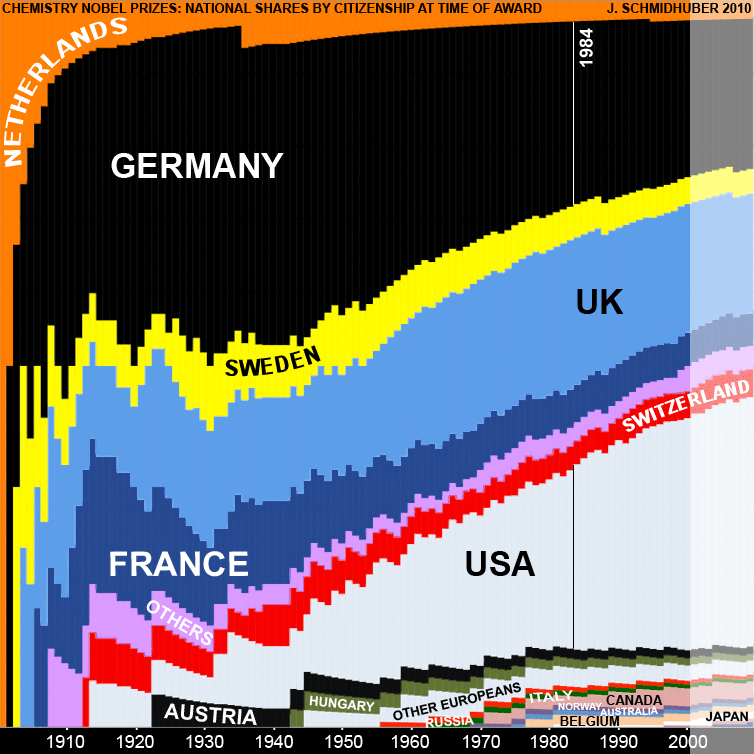}
\caption{
{\bf Chemistry} Nobel Prizes 1901-2009: 
Evolution of cumulative national shares {\bf by citizenship} at the time of the award.
The Netherlands took the lead in 1901, 
Germany caught up in 1902, 
Sweden in 1903, 
the UK in 1904; 
then Germany alone led until 1984
 (not as long as in the birth-based ranking of Figure \ref{chemnat754}); 
then the US took over.
}
\label{chem754}
\end{figure*}

\begin{figure*}[hbt]
\centering
\includegraphics[width=\textwidth,height=\textwidth]{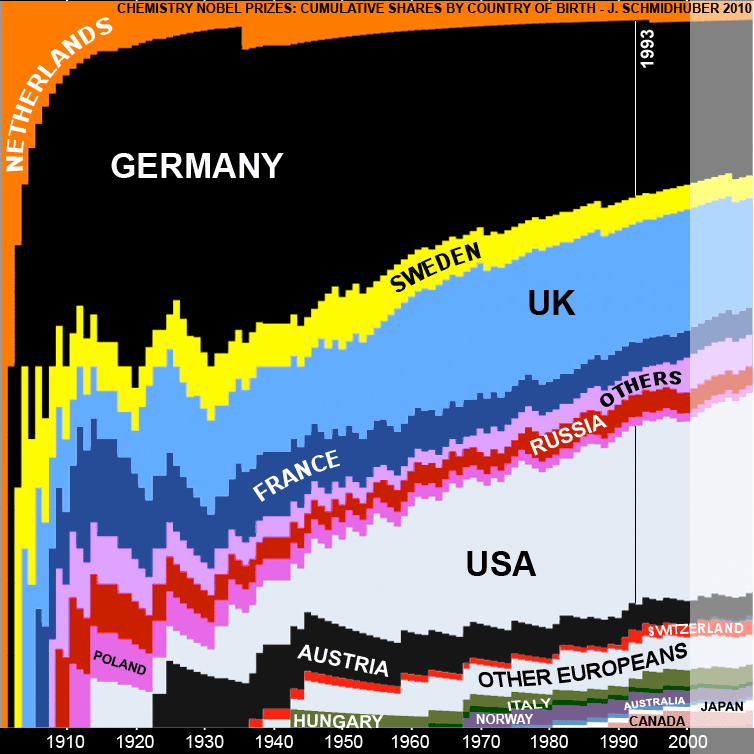}
\caption{
{\bf Chemistry} Nobel Prizes 1901-2009: 
Evolution of cumulative national shares {\bf by country of birth.}
The Netherlands took the lead in 1901, 
Germany caught up in 1902, 
Sweden in 1903, 
the UK in 1904; 
then Germany alone 
led until 1993 (longer than in the citizenship-based ranking of Figure \ref{chem754}); 
then the US took over.
}
\label{chemnat754}
\end{figure*}

\begin{figure*}[hbt]
\centering
\includegraphics[width=\textwidth,height=\textwidth]{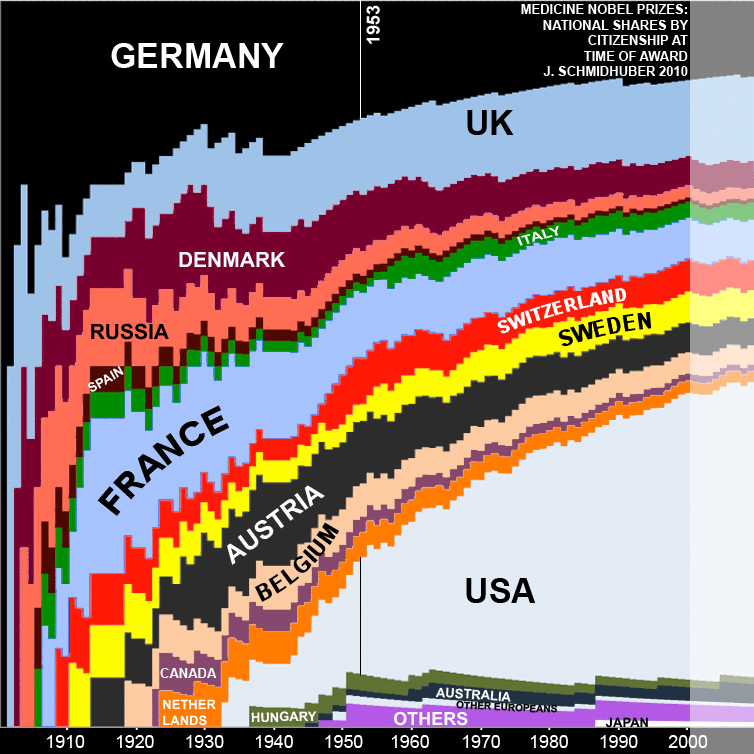}
\caption{
{\bf Medicine} Nobel Prizes 1901-2009: 
Evolution of cumulative national shares {\bf by citizenship} at the time of the award.
Germany took the lead in 1901;
the UK caught up in 1902, 
Denmark in 1903, 
Russia in 1904; 
then Germany alone led until 
1953 (not as long as in the birth-based ranking of Figure \ref{mednat754}, due to
brain drain effects); 
then the US took over.
}
\label{med754}
\end{figure*}

\begin{figure*}[hbt]
\centering
\includegraphics[width=\textwidth,height=\textwidth]{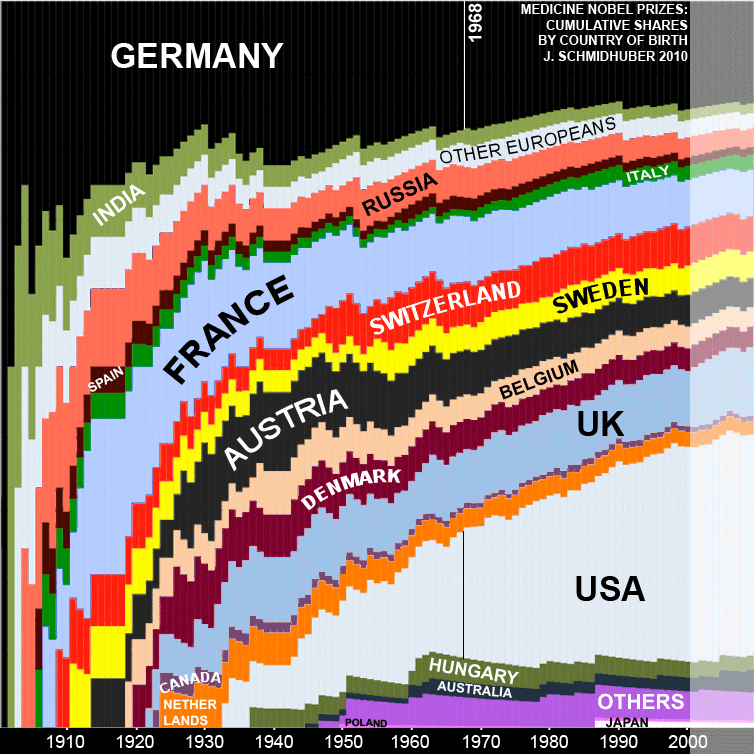}
\caption{
{\bf Medicine} Nobel Prizes 1901-2009: 
Evolution of cumulative national shares {\bf by country of birth.}
Germany took the lead in 1901;
India caught up in 1902, 
the Faroe Islands in 1903, 
Russia in 1904; 
then German-born laureates led until 1968 
(longer than in the citizenship-based ranking of Figure \ref{med754}); 
then US-born laureates took over.
}
\label{mednat754}
\end{figure*}

\begin{figure*}[hbt]
\centering
\includegraphics[width=\textwidth,height=\textwidth]{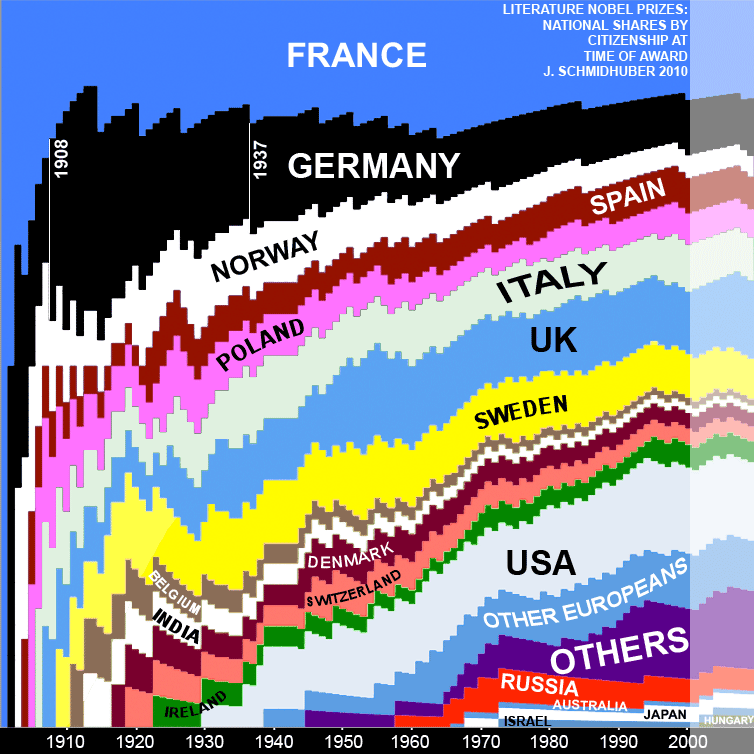}
\caption{
{\bf Literature} Nobel Prizes 1901-2009: 
Evolution of cumulative national shares {\bf by citizenship} at the time of the award.
France took the lead in 1901;
Germany caught up in 1902,
Norway in 1903,
then France alone led until 2009,
except for about three decades, 1908-1937,
when Germany was ahead.
}
\label{lit754}
\end{figure*}

\begin{figure*}[hbt]
\centering
\includegraphics[width=\textwidth,height=\textwidth]{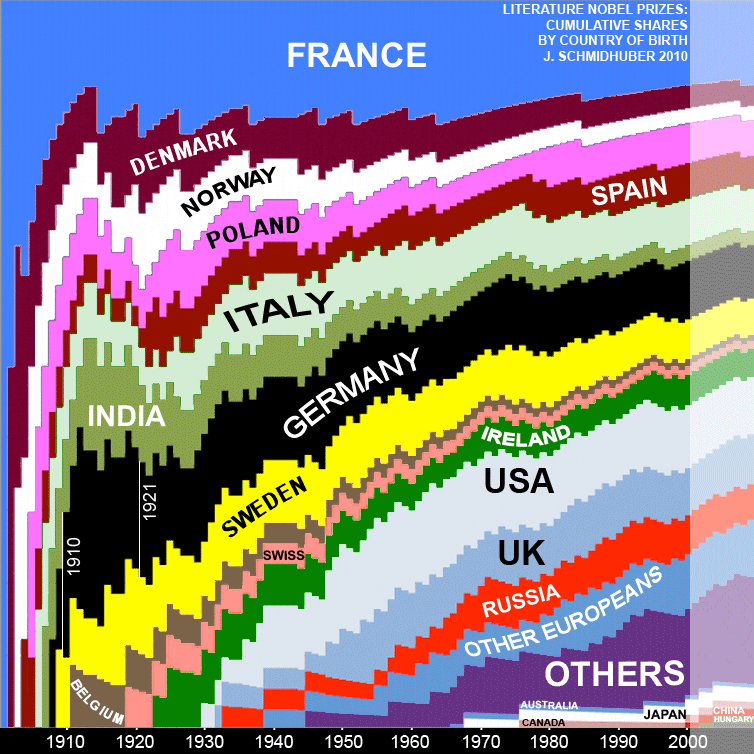}
\caption{
{\bf Literature} Nobel Prizes 1901-2009: 
Evolution of cumulative national shares {\bf by country of birth.}
France took the lead in 1901;
Denmark caught up in 1902,
Norway in 1903,
then France alone led until 2009,
except for the period 1910-1921,
when Germany was ahead.
}
\label{litnat754}
\end{figure*}

\begin{figure*}[hbt]
\centering
\includegraphics[width=\textwidth,height=\textwidth]{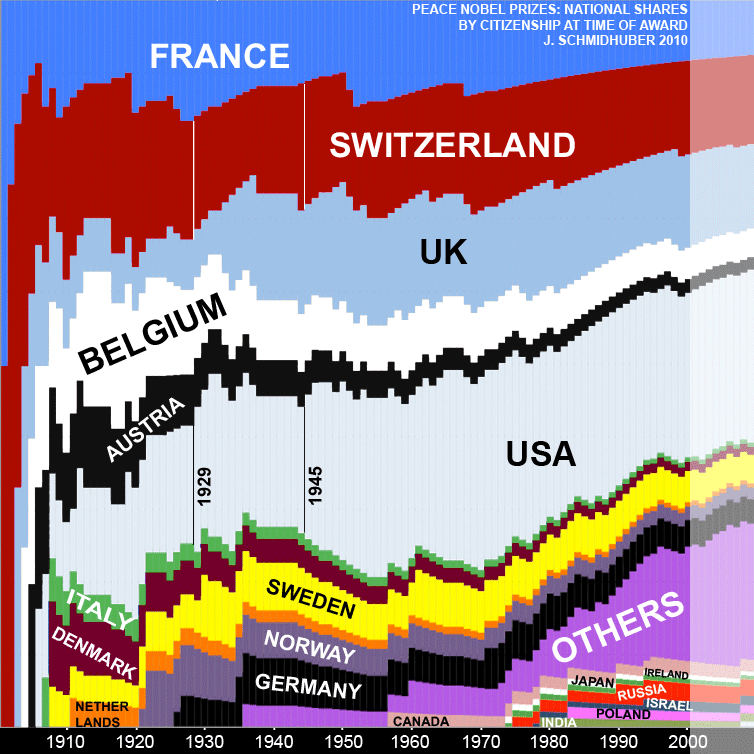}
\caption{
{\bf Peace} Nobel Prizes 1901-2009: 
Evolution of cumulative national shares {\bf by citizenship} 
(or home of organization) at the time of the award.
France and Switzerland shared the lead in 1901,
Switzerland then led alone until 1909
(when it shared the lead with France and Belgium),
then led alone until 1913,
then again shared the lead with Belgium until 1917,
then led alone until 1925, then shared the
lead with the US until 1929 (since 1927 also with the UK).
Since 1929 the US has led alone, 
except for 1944-1945 when Switzerland briefly caught up again.
}
\label{pea754}
\end{figure*}

{\bf Rankings Based on Birthplace.}
Graphs based on the laureates' countries of birth are visibly different, since
many laureates were not born in the country whose citizenship
they had when they received their prize. 
US-born laureates started to lead 
 the birth-based total Nobel Prize count in 1965 (Figure \ref{allnat754}),
roughly a decade after the corresponding citizenship-based date
reflecting earlier brain gain.
In the sciences (Figure \ref{scinat754}), ignoring initial high-variance fluctuations, 
native Germans led until 1975:
until 1993 in chemistry (Figure \ref{chemnat754}), 
until 1977 in physics (Figure \ref{physnat754}; note the particularly prominent
brain drain-caused difference to the citizenship-based ranking of Figure \ref{phys754}), 
until 1968 in medicine (Figure \ref{mednat754}).
Since 1993, all sciences have been led by US-born laureates, while
the literature count almost always was led by
French-born writers (Figure \ref{litnat754}).
Total birth-based Nobel Prize counts at the end of the century: 
US 99.58, Germany 62.58, UK 50.33, France 30.5.
In the sciences: US 74.08, Germany 51.08, UK 34.33, France 14.5.

The most obvious {\bf geographic shift} was the decline of 
Germany in the wake of World War II, and the symmetric rise of the US,
especially in the sciences (Figure \ref{scibothusger754}). 
Until shortly after the war, Germany still boasted more
Nobel Prizes (by citizenship and by birth)
than the Allied Powers  UK, US, and USSR combined 
(except for one year  1904-05). 
During this period, its share 
actually slightly profited from 
brain gain, especially in
chemistry (Figures \ref{chemnat754} \& \ref{chem754}) and literature
(Figures \ref{litnat754} \& \ref{lit754}),
not yet suffering from brain drain.
Then the picture quickly changed, as
English-speaking nations increased their share
at the expense of German-speaking and other continental nations
 (Figure \ref{scilang754}).
Asian nations also have increased their share.
As of 2009, Nobel Prize counts of major players by citizenship
are: EU $>$270, USA $\sim$150, Asia $>$30. Extrapolating
current trends, the European share may fall below 50\%
within a few decades.

{\bf Laureates per Prize.}
In the beginning of the century most laureates got a full prize;
in the end most got just a fraction thereof. This 
{\em laureate inflation} accelerated in the century's second half, when
US and UK were particularly successful.
In the 21st century this trend remains unbroken, reflected by the 
declining average prize fraction per laureate (currently about
 0.55 / 0.65 / 0.72 prizes per US / UK / German laureate). Roughly
1 in 4 (mostly  younger) US laureates, 1 in 3 UK laureates, and 1 in 2  (mostly older) German
laureates got a full prize. In the sciences, the prizes per laureate ratio
shrank even more rapidly (currently about
 0.49 / 0.69  prizes per US / German laureate;
Figure \ref{scipriperlau754}).
 Counting just 
laureates instead of their prizes  would exhibit a strong
bias towards more recent decades.

{\bf Per Capita Rankings.}
The population of the US grew from 77m in 1901 to 309m in 2009;
Germany's from 56m to 82m;
the UK's from 38m to 61m;
Switzerland's from 3.3m to 7.5m.
Since nations have grown at varying speeds, historic census data
should be taken into account  to create proper per capita measures and graphs.
This was not done here. However, 
to obtain crude {\em approximate} per capita rankings, 
one could naively divide each nation's sum of Nobel Prizes 
by its current population. 
Considering only nations whose citizens collected Nobel Prizes on a regular basis,
the ranking is led by Switzerland,
with roughly 3 Nobel Prizes per million capita (NPpmc),
followed by Sweden (nearly 2 NPpmc), 
Denmark (nearly 1.5 NPpmc), 
Austria (over 1 NPpmc), 
and the UK (about 1 NPpmc). (We
ignore statistical outliers St. Lucia and Iceland, 
each with  1.0 prizes for one single laureate,
according to the Nobel Foundation).

{\bf Individuals etc.}
The most successful Nobel Prize-winning entity
so far was the {\em International Committee of the Red Cross} 
(Switzerland, 2.5 prizes for peace).
The most successful individual 
 was L. Pauling (US, 2.0 prizes: 1.0 for peace, 1.0 for chemistry).
The top science laureates were 
M. Curie (France, 1.25 prizes: 1.0 for chemistry, 0.25 for physics) 
and F. Sanger (UK, 1.25 for chemistry).
More than 200 of over 700 laureates got exactly 1.0 prize.
The only double laureate with less than 1.0 was J. Bardeen (US, 2/3 prizes: twice 1/3 for physics).
The most successful family were the Curies
(2.5 prizes: M. Curie's 0.25 for physics, the rest in chemistry: 
1.0 for herself, 0.25 for her husband; 0.5 for I. Joliot-Curie, 0.5 for {\em her} husband).

{\bf Universities.}
Originally we intended to plot evolving Nobel Prize shares of  universities as well.
To avoid a misleading university ranking, however,
we refrained from doing this---the Nobel Foundation only lists affiliations of laureates at the 
moment they received their award, although their prize-winning breakthroughs 
were often achieved elsewhere {\em before} the listed university hired them.
Additional research is necessary to create fair university rankings 
taking into account where the distinguished work really took place,
and where the laureates received which part of their education.

{\bf Summary.} 
We discussed the growing number of laureates per Nobel Prize, and
traced patterns of brain drain and brain gain in the 20th century
 by comparing the temporal evolution of national Nobel Prize shares 
by country of birth and by citizenship.

{\bf Reference:} http://nobelprize.org, retrieved March 2010.
(For each laureate this web site used to provide prize fraction, 
nationalities, and country of birth where different from country of citizenship.)

{\bf HTML version} of this paper: {\tt http://www.idsia.ch/\~{ }juergen/nobelshare.html}


\begin{figure*}[hbt]
\centering
\includegraphics[width=\textwidth,height=\textwidth]{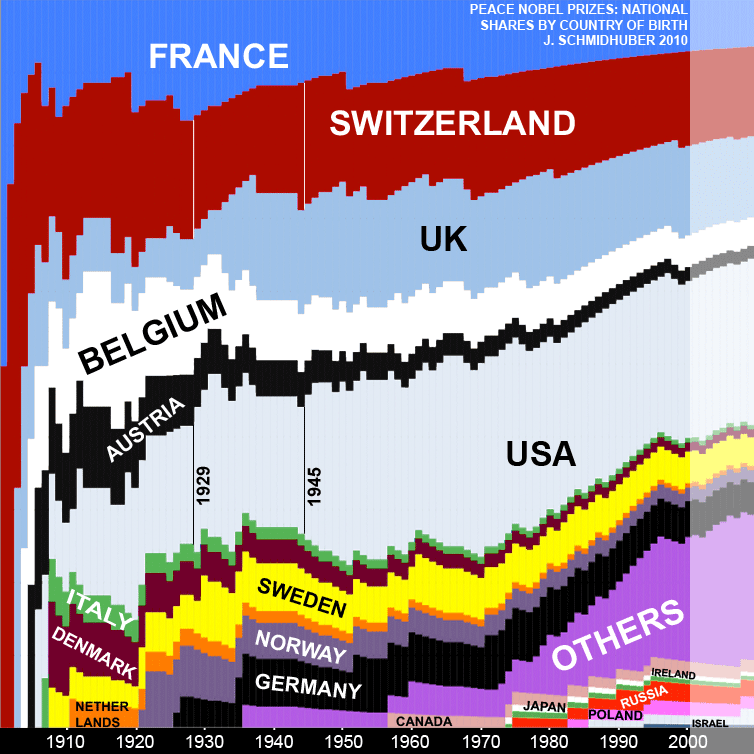}
\caption{
{\bf Peace} Nobel Prizes 1901-2009: 
Evolution of cumulative national shares {\bf by country of birth} (or 
nation in which organization was founded).
Here the differences to the citizenship-based ranking (Figure \ref{pea754}) are minor.
France and Switzerland shared the lead in 1901,
Switzerland then led alone until 1909
(when it shared the lead with France and Belgium),
then led alone until 1913,
then again shared the lead with Belgium until 1917,
then led alone until 1925, then shared the
lead with the US until 1929 (since 1927 also with the UK).
Since 1929 the US has led alone, 
except for 1944-1945 when Switzerland briefly caught up again.
}
\label{peanat754}
\end{figure*}

\begin{figure*}[hbt]
\centering
\includegraphics[width=\textwidth,height=\textwidth]{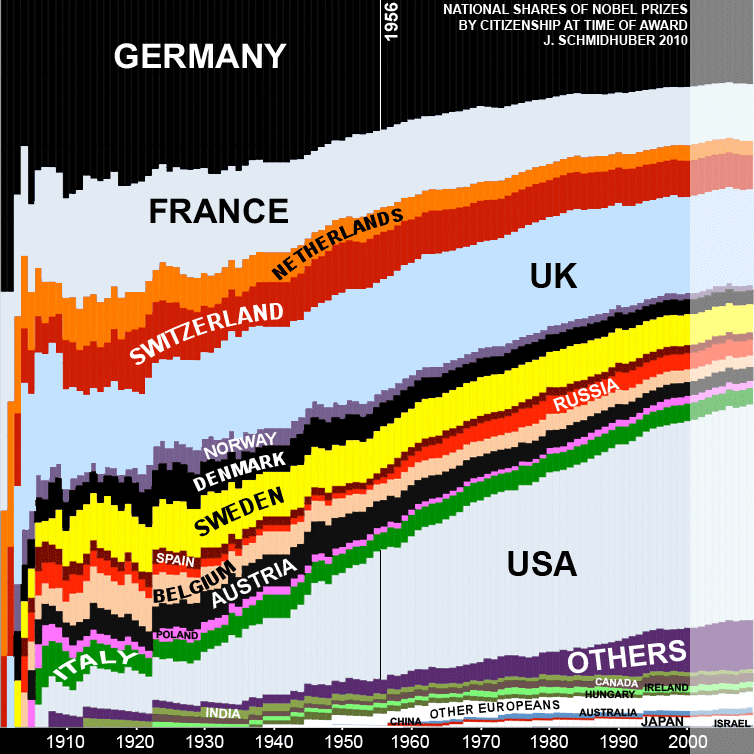}
\caption{
{\bf All} Nobel Prizes 1901-2009: 
Evolution of cumulative national shares {\bf by citizenship} at the time of the award.
From 1901-1956,
Germany led the total Nobel Prize count (sharing the lead with the UK for one
year 1904-05). Then the US surged ahead. 
}
\label{all754}
\end{figure*}

\begin{figure*}[hbt]
\centering
\includegraphics[width=\textwidth,height=\textwidth]{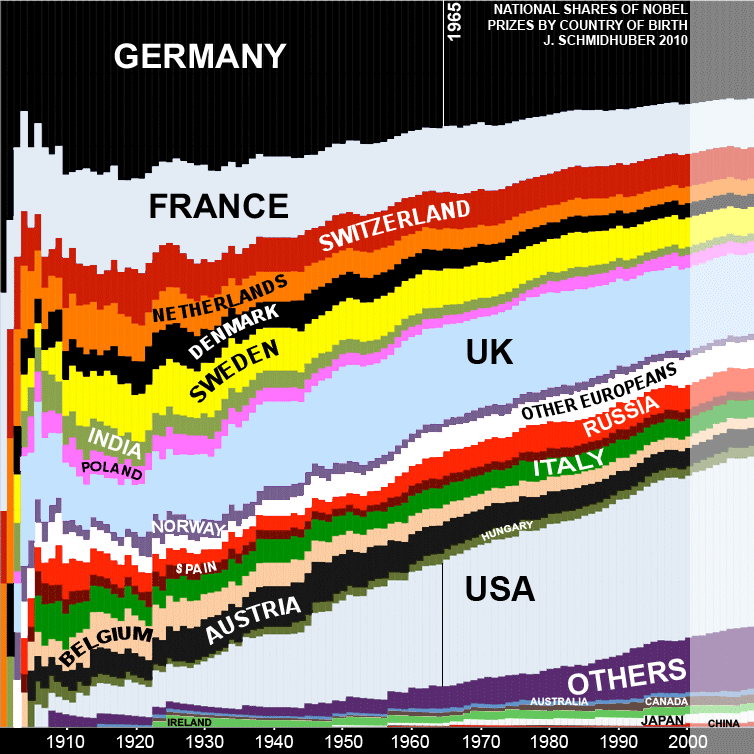}
\caption{
{\bf All} Nobel Prizes 1901-2009: 
Evolution of cumulative national shares {\bf by country of birth.}
German-born laureates led from 1901-1965 (sharing the lead with UK-born for one
year 1904-05). Then US-born laureates took over.
}
\label{allnat754}
\end{figure*}

\begin{figure*}[hbt]
\centering
\includegraphics[width=\textwidth,height=\textwidth]{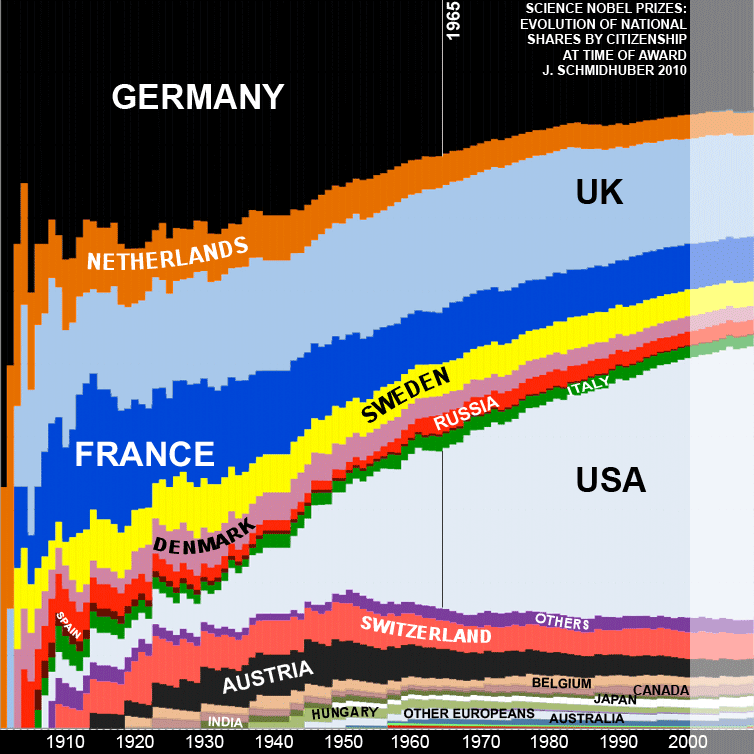}
\caption{
{\bf Science} Nobel Prizes 1901-2009: 
Evolution of cumulative national shares {\bf by citizenship} at the time of the award.
Germany led from 1901-1964 (sharing the lead with the UK for one
year 1904-05). Then the US took over. 
Note  the substantial differences 
to the birth-based ranking of Figure \ref{scinat754},
reflecting brain drain from Germany and other countries to the US (and also the  UK).
}
\label{sci754}
\end{figure*}

\begin{figure*}[hbt]
\centering
\includegraphics[width=\textwidth,height=\textwidth]{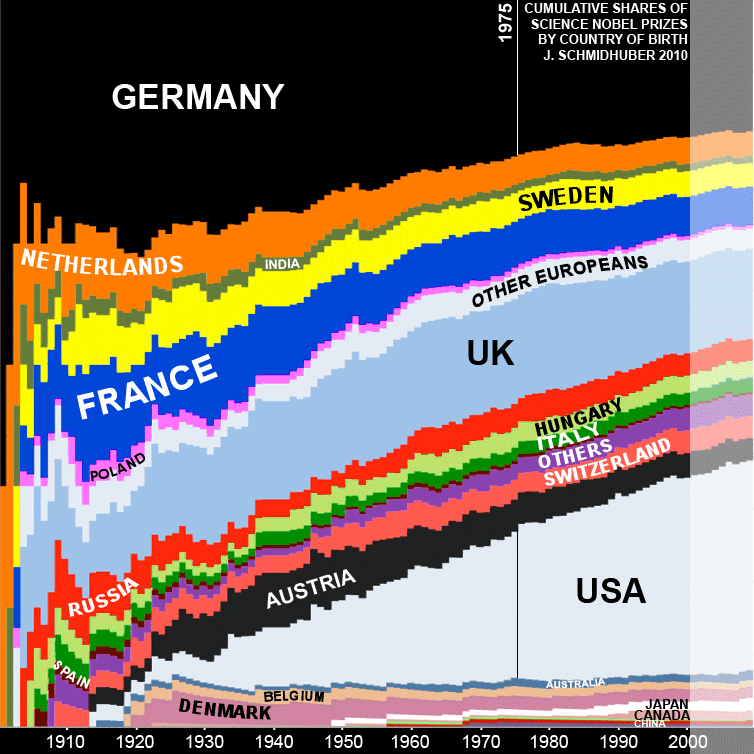}
\caption{
{\bf Science} Nobel Prizes 1901-2009: 
Evolution of cumulative national shares {\bf by country of birth.}
German-born laureates were ahead during the first three quarters of the century;
in 1976 US-born laureates took over. 
Note  the differences 
to the citizenship-based ranking of Figure \ref{sci754},
reflecting brain drain to the US (and also the  UK).
}
\label{scinat754}
\end{figure*}

\begin{figure*}[hbt]
\centering
\includegraphics[width=\textwidth,height=\textwidth]{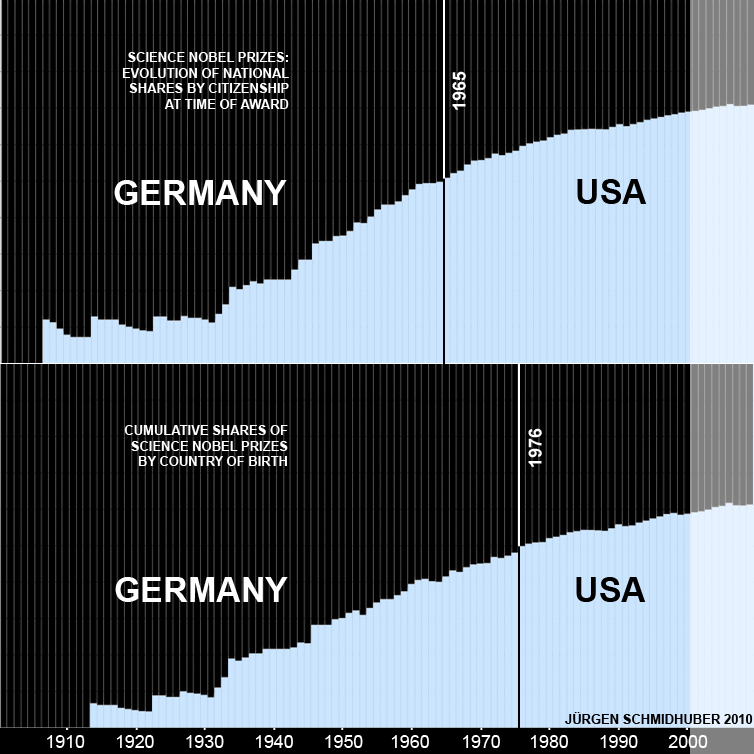}
\caption{
Science Nobel Prizes 1901-2009: 
Evolution of cumulative national shares {\bf by citizenship} 
at the time of the award (top), and
{\bf by country of birth} (bottom),
considering only the two most successful nations,
to focus on this particularly prominent geographic shift
during the 20th century. 
Note the differences between the  birth-based and the citizenship-based ranking,
due to brain drain from Germany and other countries to the US:
In the last third of the century, US citizens took over; 
in the last quarter of the century, US-born laureates took over. 
}
\label{scibothusger754}
\end{figure*}

\begin{figure*}[hbt]
\centering
\includegraphics[width=\textwidth,height=\textwidth]{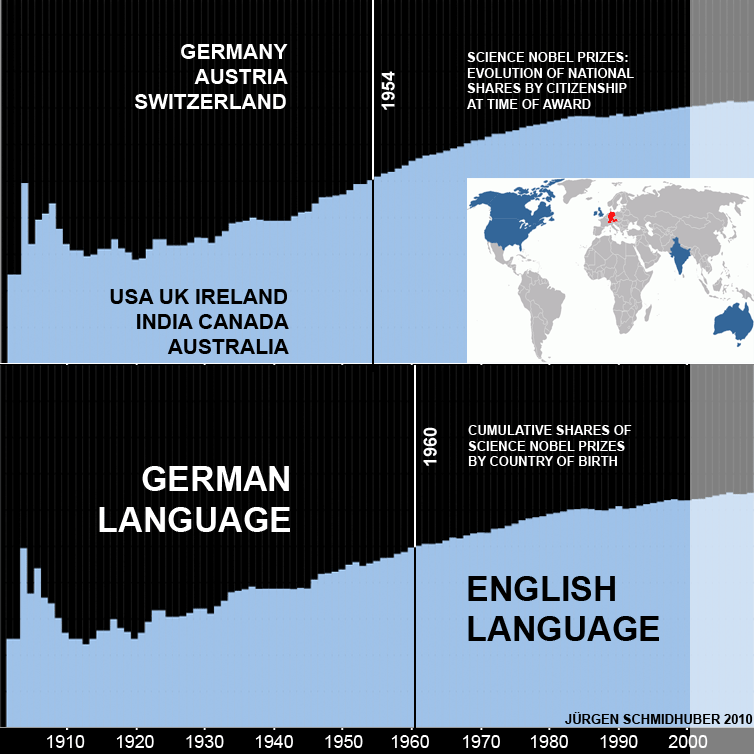}
\caption{
Science Nobel Prizes 1901-2009 {\bf by language}, illustrating the global shift from 
German to English as main language of science. Again we plot the
evolution of cumulative shares by citizenship
at the time of the award (top), and
by country of birth (bottom),
considering only nations whose dominant language is English
(we include UK, USA, Ireland, Australia, Canada, India---blue in the map)
or German (Germany, Austria, Switzerland---red in the map).
Laureates of mostly German-speaking nations were ahead until 1954
in the citizenship-based ranking, and until 1960 in the birth-based ranking;
then English took over for good.
}
\label{scilang754}
\end{figure*}

\begin{figure*}[hbt]
\centering
\includegraphics[width=10cm]{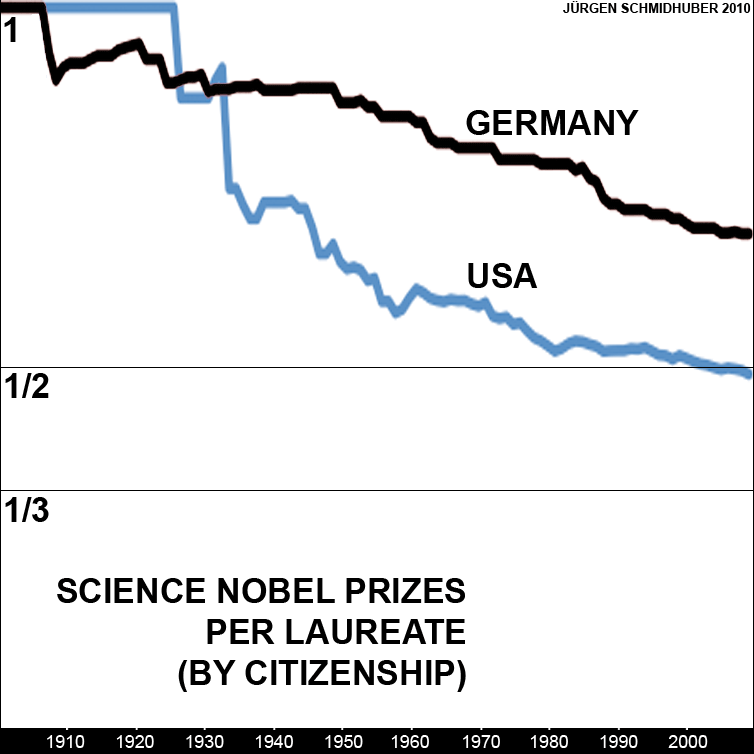}
\caption{
Science Nobel Prizes per science laureate by citizenship 1901-2009:
In the beginning of the century most laureates got a full prize;
in the end most got just a fraction thereof. That's why
Germany's mostly older laureates on average got larger
shares than the mostly younger ones of the US.
}
\label{scipriperlau754}
\end{figure*}

\bibliography{bib}
\bibliographystyle{alpha}
\end{document}